# Some New Research Trends in Wirelessly Powered Communications

Kaibin Huang[1], Caijun Zhong[2] and Guangxu Zhu[3]

## Abstract

The vision of seamlessly integrating *information transfer* (IT) and microwave based *power transfer* (PT) in the same system has led to the emergence of a new research area, called *wirelessly powered communications* (WPC). Extensive research has been conducted on developing WPC theory and techniques, building on the extremely rich wireless communications literature covering diversified topics such as transmissions, resource allocations, medium access control, and network protocols and architectures. Despite these research efforts, transforming WPC from theory to practice still faces many unsolved problems concerning issues such as mobile complexity, power transfer efficiency, and safety. Furthermore, the fundamental limits of WPC remain largely unknown. Recent attempts to address these open issues have resulted in the emergence of numerous new research trends in the WPC area. A few promising trends are introduced in this article. From the practical perspective, the use of backscatter antennas can support WPC for low-complexity passive devices, the design of spiky waveforms can improve the PT efficiency, and analog spatial decoupling is proposed for solving the *PT-IT near-far problem* in WPC. From the theoretic perspective, the fundamental limit of WPC can be quantified by leveraging recent results on *super-directivity* and the limit can be improved by the deployment of large-scale distributed antenna arrays. Specific research problems along these trends are discussed, whose solutions can lead to significant advancements in WPC.

## I. Introduction

The idea of wireless power transfer using *radio-frequency* (RF) waves can be traced back to Nikola Tesla's ambitious Wardencliffe tower project more than 100 years ago. Despite its initial failure, driven by the advancements in microwave technologies, recent years have witnessed a tremendous development of *microwave power transfer* (MPT) and its successful adoption in commercial systems such as RFID tags [1]. The ability of cutting the "last wires", i.e., the cables for recharging mobile devices, in addition to the appealing characteristics of fully controllable energy output, has positioned MPT as a promising solution for powering energy-constrained and low-rate wireless networks.

The integration of MPT and wireless communications has opened up a new and exciting research area, called *wirelessly powered communications* (WPC) [2, 3]. Due to high attenuation of wireless propagation, WPC has thus far been primarily used in short-range wireless networks with low-power nodes such as RFID and sensors. However, with the evolution of cellular networks into 5G, the newly emerging wireless technologies such as *small cell network*s and *massive MIMO* can significantly reduce the propagation loss by shortening communication distances and signal dispersion. In the meantime, the power consumption of wireless devices is expected to decrease continuously. Therefore, it is predicted that WPC will become an indispensable building block for many commercial and industry wireless systems in the future, including the upcoming *Internet of Things* (IoT) systems, wireless sensor networks, RFID networks and small cell networks [2, 3]. Presently WPC is one of the most active areas in wireless communications with extensive research being conducted on the WPC's fundamental limits, transmission techniques, resource allocation, network architectures and medium access control. With the rapid advancements in enabling technologies, it is believed that WPC has the potential of powering devices larger than low-power sensors and tags, e.g., wearable computing devices or even smartphones [3]. This has shifted the paradigm of WPC research from being energy-efficiency centric to the present one based on more complex metrics such as energy-rate tradeoff. Latest advancements on these topics have been surveyed in a recent special issue of IEEE Communications Magazine [4].

---

[1] K. Huang is with the Dept. of Electrical and Electronic Engineering at The University of Hong Kong, Hong Kong. Email: huangkb@eee.hku.hk.
[2] C. Zhong is with the Dept. of Information Science and Electronic Engineering, Zhejiang University, China. Email: caijunzhong@zju.edu.cn.
[3] G. Zhu is with the Dept. of Electrical and Electronic Engineering at The University of Hong Kong, Hong Kong. Email: gxzhu@eee.hku.hk.

In the literature, there exist two typical system configurations for implementing WPC: one is referred to as the *simultaneous wireless information and power transfer* (SWIPT) *system* [5] and the other *PB-assisted WPC system* [6], which are illustrated in Fig. 1. In a SWIPT system shown in Fig. 1(a), a *base station* (BS) transmits a modulated wave from which a target mobile harvests energy and demodulates information bits. Due to the usually long propagation distance (e.g., more than hundred meters) from the BS to a mobile, a SWIPT system has a low *power transfer* (PT) efficiency and is suitable only for powering low-power devices such as sensors and RFID. The drawback can be overcome by deploying a *power beacon* (PB) dedicated for PT, resulting in a PB-assisted WPC system shown in Fig. 1(b). PBs have low complexity and require only connections to the electricity grid and hence can be deployed with a high density that reduces the PT range and loss. A PB may be integrated with a femto-cell BS with a backhaul link to support short-range SWIPT to mobiles (see Fig. 1(b)).

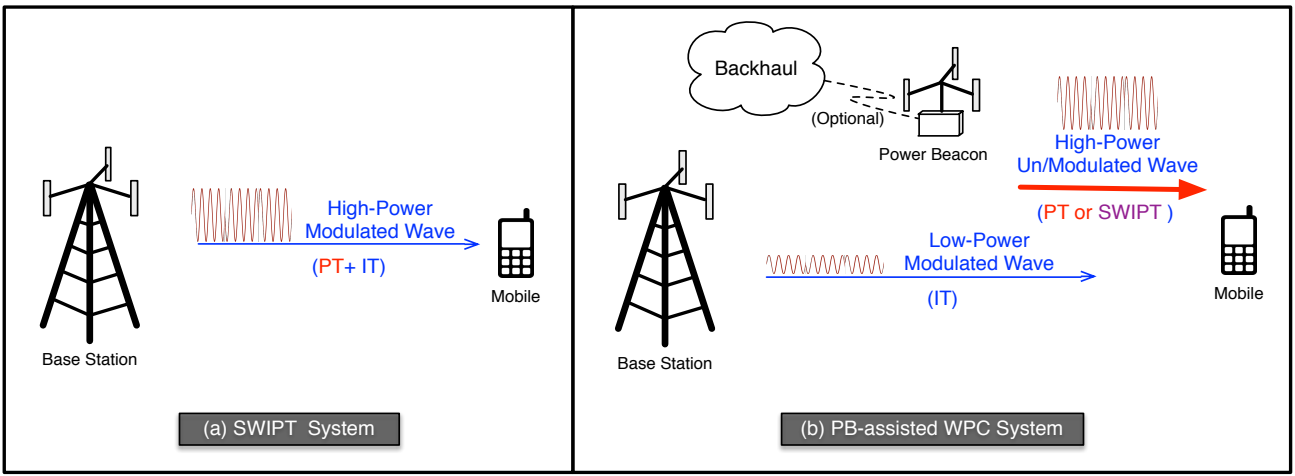

Fig. 1. Two typical WPC system configurations: (a) a SWIPT system and (b) a PB-assisted WPC system.

Despite rapid progress from the theoretical perspective, the realization of SWIPT still faces many challenges. The attempts by researchers on bridging the SWIPT theory and practice have led to the recent emergence of new research trends addressing different aspects of the technology such as efficiency, complexity and safety. In this article, we identify several promising trends as summarized below and discuss research opportunities in these directions.

- **Backscatter WPC**: Equipped with backscatter antennas, low-complexity wireless devices without active RF components, such as RFID, are able to receive energy from and transmit data to BSs, called *backscatter WPC*. In **Section II**, backscatter (enabled) WPC are introduced and different relevant research opportunities are discussed including full duplexing, network architectures, scheduling and the integration of backscatter WPC and retrodirective beamforming. Both SWIPT and PB-assisted WPC systems are considered in the discussion.

- **Waveform designs for SWIPT**: As introduced in **Section III**, the recent experimental finding of a relation between the signal spikiness and energy-harvesting efficiency opens up the possibility of improving the PT efficiency of a SWIPT system by spiky waveform designs. Nevertheless, such designs have to balance the *information-transfer* (IT) rate and PT efficiency, creating a number of interesting research opportunities. In Section III, several are discussed including: 1) robust waveform design for maintaining signal spikiness in multi-path propagation, 2) waveform design under the bandwidth and hardware constraints, and 3) waveform design that optimizes the tradeoff between the IT rate and PT efficiency, called the *energy-rate tradeoff*. The discussion assumes the SWIPT system.

- **Super-directive SWIPT**: A fundamental question for SWIPT is "Can the IT rate and PT efficiency keep growing by packing more antennas into a transmit array?" The answer is *negative* as revealed by recent research on *super-directivity* as discussed in **Section IV**. Building on super-directivity results, without limiting the number of transmit antennas, it is possible to derive a fundamental energy-rate tradeoff for SWIPT. Several relevant research opportunities are discussed in Section IV, targeting the SWIPT system.

- **SWIPT using ubiquitous arrays**: The practical limitations of a super-directive array can be overcome by distributed deployment of the antennas. The resultant large-scale distributed array, call a *ubiquitous array*, not only supports efficient SWIPT but also provides other advantages such as enabling blind channel estimation and guaranteeing lines-of-sight in the presence of scattering. The relevant discussion is presented in **Section V** targeting the SWIPT system.

- **Analog spatial decoupling of PT and IT for WPC**: In a PB-assisted WPC system, the powers of PT and pure IT signals received at a mobile can differ by many orders of magnitude. Quantizing the mixed signals may cause the IT signal to be completely corrupted by quantization noise. This motivates the need of decoupling the two types of signals in the analog domain by exploiting their spatial separation. Designing analog decoupling circuits using simple components such as phase shifters is challenging and the theme of **Section VI**.

## II. WPC Over Backscattering Channels

**a) Backscatter Communications**

In a backscatter communication system illustrated in Fig. 2, a mobile device not only harvests energy from an incident sinusoidal wave radiated by a transmitter but also modulates and reflects a fraction of the wave back to the transmitter. The wave reflection is due to an intentional mismatch between the antenna and load impedance. Varying the load impedance causes the complex scatter coefficient to vary following a random sequence that modulates the reflected wave with information bits (see Fig. 2). Backscatter communication allows a transmitter to power a passive receiver that uses no active RF components. This unique type of communication has received considerable interests in recent years, mainly due to its successful implementation in passive RFID systems and the potential use in other wireless devices with low power and small form factors such as sensors [7].

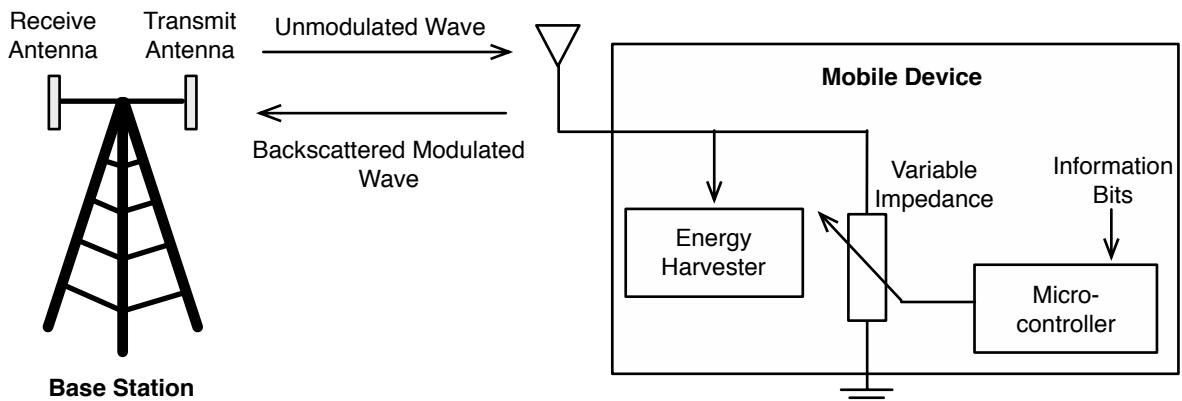

Fig. 2. Backscatter communication systems.

Two key features of backscatter communication systems differentiate them from conventional ones.

- **Dyadic MIMO channels**. The implementation of backscatter in a MIMO system creates a special class of MIMO channels, called *dyadic MIMO channels,* that capture the composite fading in the forward and backward channels [7]. Specifically, the unmodulated wave radiated by the transmitter propagates through the forward MIMO channel, modulated and scattered by the receiver and last propagates through the backward MIMO channel to arrive at the transmitter. As a result, the effective dyadic channel can be written as the product of two matrices $\mathbf{H}_{UL} \times \mathbf{D}_{DL}$ where $\mathbf{H}_{UL}$ is the uplink channel matrix and $\mathbf{D}_{DL}$ is a diagonal matrix computed from the downlink channel matrix. The structure of the dyadic MIMO channels results in new channel models unlike the conventional ones (e.g., i.i.d. Rayleigh fading) and affects the channel capacity and designs of transmission schemes such as space-time coding and diversity-multiplexing tradeoffs [7].

- **Energy-rate tradeoffs**. There exist two energy-rate tradeoffs in a backscatter communication system. The first arises from the duty cycle of the passive receiver defined as the fraction of time the receiver transmits information by backscattering. The remaining time is spent on idling and energy harvesting. Note that backscattering reduces harvested energy as part of it is scattered. Then a tradeoff exists where increasing the duty cycle reduces harvested energy but increases the information rate and vice versa. The second tradeoff results from the choice of modulation constellation size. The growth of the number of bits per symbol requires a larger constellation size that increases the average reflection coefficient and as a result reduces the average harvested energy, yielding a corresponding energy-rate tradeoff.

Backscatter communication provides a useful basic technique for designing low-power and lower-complexity mobile devices and building novel WPC systems for serving such devices.

b) **Backscatter Enabled WPC and Research Opportunities**

WPC building on the principle of backscatter communication, called *backscatter WPC*, is a promising research direction. Some initial research in this direction has been pursued in [8] where a channel estimation technique customized for backscatter enabled MPT is proposed. However, the area of backscatter enabled WPC is largely uncharted. Several potential research opportunities are described as follows.

- **Full-duplex backscatter WPC**. A conventional backscatter communication system (see Fig. 2) supports simultaneous downlink transmission of an *unmodulated* sinusoidal wave and uplink transmission of modulated scattered signal. One interesting research direction is to also modulate the said sinusoidal wave with information, supporting full-duplex IT combined with downlink PT. The key challenge lies in decoding the signal at the BS that has been *double* modulated with downlink and uplink information bits. However, by exploiting the prior knowledge of downlink information, it is possible for the BS to retrieve the uplink information. This motivates the design of special modulation and precoding schemes to facilitate the decoding of double modulated signal by leveraging well-known techniques such as dirty paper coding or Tomlinson-Harashima precoding.

- **Backscatter enabled WPC networks**. The implementation of backscatter at mobile devices reduces their power consumption and facilitates MPT and thus finds applications in upcoming IoT or sensor networks with MPT. In PB assisted WPC networks, mobile devices can harvest energy from part of the wave transmitted by PBs and modulate and scatter the remaining to BSs. For designing such networks, the network throughput can be maximized by optimizing different configurations including the parameters of mobile backscatter antennas (e.g., reflection coefficients, constellation size and duty cycle) and the mentioned energy-rate tradeoffs for backscatter communications.

- **Backscatter enabled retrodirective beamforming for WPC**. Retrodirective beamforming is a simple technology for a multi-antenna BS to automatically steer a radio beam towards a target user based on the signal transmitted by the user and received at the BS. The beam is formed by conjugating the phase of the signal received at each antenna and retransmitting the signal after amplification. Thus, retrodirective beamforming requires neither knowledge of the user location nor any complex beamforming algorithm. There are different ways for realizing phase conjugation without requiring phase shifters or digital signal processing [9]. It is interesting to integrate backscatter communication and retrodirective beamforming to realize *automatic SWIPT* supporting fast beam tracking. In such a design, a BS periodically broadcasts a pilot signal (e.g., a continuous wave) and based on the backscattered signals, steers retrodirective beams towards multiple users. Due to their automatic operation, such beams are able to track the movements of users even at high speeds. Furthermore, automatic SWIPT does not require channel estimation or CSI feedback at mobiles and thus reduces their complexity. One drawback of automatic SWIPT is that the retrodirective beamforming does not provide flexibility over the beam pattern and is hence incapable of interference avoidance. Then for automatic SWIPT, spatial division multiple access has to rely on a large-scale transmit array or large user spatial separation for suppressing multiuser interference.

- **Scheduling for backscatter WPC**. The powering of mobile uplink transmission by downlink MPT, e.g., for the case of backscatter communication, causes double (uplink + downlink) attenuation, resulting in a severe near-far problem in WPC systems [2]. On one hand, the mobile device located far-away from the BS harvests less energy. On the other hand, it requires more energy for uplink IT to compensate for severe path loss. As such, mobile devices near the BS are more likely to be served if the objective is simply to maximize the spectral or energy efficiency, leaving no or very few transmission opportunities for the far-away mobile devices. Therefore, how to design a sophisticated multi-user scheduling algorithm guaranteeing fairness is an important issue to be addressed.

## III. Waveform Design for SWIPT

A typical MPT system is illustrated in Fig. 3, the RF wave radiated by the PB is captured by the antenna at the energy receiver, and fed into a rectifying circuit. The circuit usually consists of a single diode and low pass filter (LPF), which converts the RF signal back to the DC energy for storage. To implement a MPT system, one of the most critical issues to address is how to improve the RF-to-DC conversion efficiency. Thus far, major efforts have been devoted to the design of efficient *rectennas*, each integrates an antenna and a rectifier. However, the overall RF-to-DC conversion efficiency not only depends on the efficiency of rectenna, but is also closely related to the input waveform as discussed shortly.

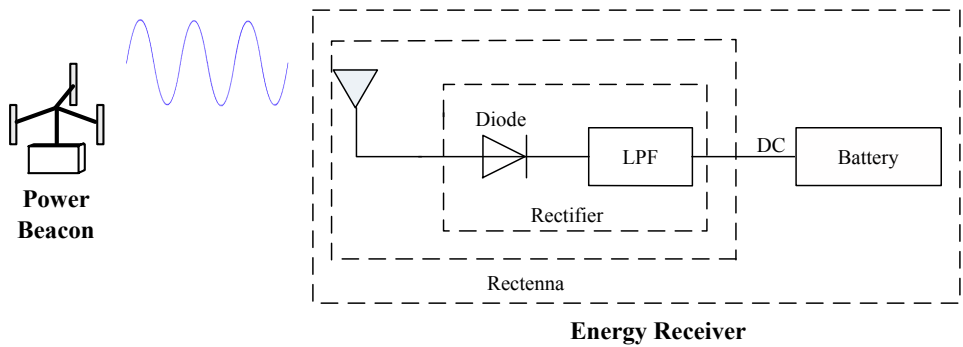

Fig. 3. Microwave power transfer system.

Upon receiving a single carrier continuous-wave (CW) signal, ideally, the output DC of the rectenna is proportional to the input power because of the dominant second-order behavior of the diode, an assumption widely adopted in the SWIPT literature. However, in practice, due to the nonlinear behavior of the diode, the output DC not only depends on the average input power, but also relies on the fourth order of the input signal, i.e.,

$$y(t) = k_2 x^2(t) + k_4 x^4(t)$$

where $x(t)$ is the input signal, $y(t)$ is the output current, and $k_2$ and $k_4$ are diode-specific constants. This important theoretical foundation has sparked an upsurge of research interests in energy efficient waveform design. In particular, it has been recently demonstrated through experiments that using multisine signal yields higher output DC compared to the single-carrier CW signal with the same average transmit power. This has led to the conclusion that *spiky* signals, i.e., signals with high *peak-to-average power ratio* (PAPR), tend to achieve a higher energy conversion efficiency [10].

The exciting development in the MPT community has been largely overlooked in the SWIPT community. Only in a recent work [11], building upon these latest results, a step forward was made: optimize the design of multisine waveforms with multiple transmit antennas. Since the waveform design for MPT is a newly emerged topic, there are still many important issues to address. In particular, we identify the following potential research opportunities.

- **Robust waveform design for MPT**: For the waveform design of multisine unmodulated signals for MPT, phase alignment of individual subcarriers is essential for obtaining a highly spiky signal. In addition, subcarrier amplitudes and frequency separation can be played to further improve the performance. When the generated RF energy signals propagate over the wireless medium, random phase and amplitude distortions will corrupt the RF signal due to multi-path fading. Therefore, it is imperative to compensate for such random distortion at the PB, which, however, requires accurate CSI. In practice, obtaining accurate CSI for MPT systems may be difficult and incurs extra overhead. Therefore, how to design the signal waveform based on only partial CSI with a guaranteed energy efficiency is a problem of practical interest.

- **Bandwidth constrained waveform design**: Consider a PB-assisted WPC system. The single carrier CW signal transmitted by the PB effectively uses zero bandwidth, while the spiky signals, e.g., multisine, OFDM and UWB signals, tend to occupy nonzero or even large bandwidth. Consequently, the improvement on the energy conversion efficiency by transmitting spiky signals comes at the price of bandwidth expansion. In practice, wide-band PB transmission may not be desirable for several reasons. First, spectrum sharing between the PT and IT cause the former to interfere with the latter. Second, wide-band PB transmission imposes extra design requirements for energy-harvesting circuits, i.e., wide-band rectennas. Therefore, it is important to study the optimal waveform design for maximizing the energy conversion efficiency under a bandwidth constraint.

- **Waveform design for optimal energy-rate tradeoff**: For SWIPT applications, it is desirable to use the signal waveform that is efficient for both the PT and IT. However, these two objectives are in conflict. On one hand, spiky signals are preferred in terms of enhancing the energy conversion efficiency. On the other hand, spiky signals encounter non-linear signal distortion due to the limited dynamic range of practical power amplifiers, which causes substantial IT rate loss. Therefore, there exists an energy-rate tradeoff as a function of signal spikiness, which is an interesting problem to investigate.

# IV. Super-Directive SWIPT

## a) Super-Directivity and SWIPT

The directivity of free-space energy beamforming determines the MPT efficiency. It is important to investigate its fundamental limit by studying the extreme of directivity as more antennas are packed into an array with a fix length or aperture area. This leads to a phenomenon called *super-directivity* that refers to directivity of a dense array much higher than a reference array (e.g., a uniform linear array with half-wavelength antenna separation), which has recently received lots of attention in the literature. In the context of SWIPT, studying super-directivity gives rise to a new and fundamental energy-rate tradeoff regulated by directivity as elaborated in the sequel.

In theory, the finite size of an array aperture cannot place any limit on the directivity and the number of data streams spatially multiplexed (or equivalently the number of spatial DoF). The reason is that an arbitrary radiation pattern can be generated by an aperture of any given size so long as the antennas packed into the aperture are sufficiently dense. However, implementing super-directivity in practice is infeasible due to several issues [13]. First, given an array enclosed in a space of fixed volume, increasing the antenna density leads to very large magnitudes and extremely fast phase variations in the excitation coefficients of the array elements. This results in large and oppositely directed currents in adjacent elements. Then realizing super-directivity requires extremely precise adjustments of antenna excitation coefficients. Second, huge current variations in a super-directive array causes a large *Q factor*, defined as the ratio between the stored energy and radiated energy. An array with a large Q factor has low efficiency because a large amount of energy is trapped in the evanescent field in the vicinity of the array instead of being radiated as electromagnetic waves. Another effect of increasing the Q factor is the reduction of the array bandwidth. Last, the very large currents in the elements of super-directive arrays cause the ohmic losses to increase dramatically, resulting in severe reduction of array efficiency. The ohmic loss can be reduced by using superconductive material. In conclusion, it is impossible to achieve unlimited directivity in practice but a fair-amount of super-directivity can be attained by optimizing transmission designs under practical constraints. Typical constraints include the accuracy of the realized beam pattern, the maximum ohmic loss, the maximum reactive (non-radiative) power and the maximum radiative power.

Consider a MIMO channel with a transmit array with unlimited elements and enclosed in a sphere with a fixed radius. The fundamental limit of the channel capacity as achieved by super-directivity is quantified in [14] under constraints on both the total radiated and reactive powers. The spatial modes used for spatial multiplexing are created by decomposing the electromagnetic field using spherical harmonic functions. Thereby, it was discovered in [14] that the maximum number of DoF is proportional to the array aperture area. Another key finding is that increasing directivity allows more modes to be excited at the price of increasing Q factor and consequently reducing the bandwidth. This implies that super-directivity increases the *spatial DoF* at the expense of the *spectral DoF*. Therefore, the information capacity gain of super-directivity is finite and may not even be large.

From the IT perspective, the results in [14] establish the relation that increasing super-directivity increases spatial DoF but reduces bandwidth, which are two key factors determining the channel capacity. From the PT perspective, increasing super-directivity suppresses propagation loss but also reduces array efficiency, which have opposite effects on the PT efficiency. Jointly consideration the above relations reveals the existence of a new and fundamental energy-rate tradeoff induced by super-directivity.

**b) Research Opportunities**

The optimization of the mentioned energy-rate tradeoff under various practical constraints introduces a new set of design challenges for highly efficient SWIPT as summarized below.

- **Super-directivity induced energy-rate tradeoff**: Quantifying the tradeoff requires the derivation of the spatial DoF, bandwidth and PT efficiency as functions of super-directivity.

- **Robust super-directive SWIPT**: In practice, inaccuracy inevitably exists in the amplitudes and phases of the excitation coefficients of a super-directive array. The array functions can be designed to reduce the sensitivity of the element values to the said inaccuracy. Alternatively, under bounded errors for the excitation coefficients, super-directive transmission algorithms can be designed to optimize the worst-case SWIPT performance.

- **End-to-end efficiency maximization for super-directive SWIPT**: The energy-rate tradeoff can be characterized by accounting for the effects of super-directivity on the transmit array efficiency and energy harvesting efficiency at the receiver. This will yield the end-to-end efficiency for implementation of super-directive SWIPT.

- **Waveform design for super-directive SWIPT**: Waveforms having large PAPRs are known to improve the energy harvesting efficiency (see Section III). However, such waveforms require relatively large bandwidth. Jointly considering this relation and that between super-directivity and bandwidth makes it necessary to design optimal waveforms for super-directive SWIPT.

## V. SWIPT Using Ubiquitous Arrays

**a) Ubiquitous Arrays and SWIPT**

The merger of latest technologies including C-RAN, super-dense small cells and massive MIMO suggests that in a future system outdoor mobiles can be surrounded by a distributed array having thousands of elements or even more as illustrated in Fig. 4 (a). Such an array with an unprecedented area, density and scale is named a *ubiquitous array* (UA) [15]. With the advancements in printable antennas, a UA can be also implemented in the indoor environment by installing large antenna panels in the ceilings, walls and grounds as shown in Fig. 4 (b). The deployment of UAs can dramatically increase the PT efficiency and IT capacity. This will also lead to sharp reduction of transmission power, thereby supporting green communication and alleviating the safety issue for MPT.

Exploiting its large geometry, a UA can be used to shape the electromagnetic field such that radiated power is concentrated in a small region centered at the targeted user, or equivalently forming an *energy hotspot*. In contrast, a large-scale array with elements contained in a small space, namely a super-directive array discussed earlier, forms a sharp beam in the direction of the user. A UA has several advantages over a super-directive array. First, with elements distributed over a large space, a UA does not encounter the practical issues faced in implementing super-directivity e.g., reduced bandwidth and low array efficiency caused by high Q factor and required high-precision control. Second, the sharp beam formed by the super-directive array can be easily blocked by an object e.g., building, furniture or human while it is unlikely to block the lines-of-sight of all UA elements. Thus, a UA can ensure efficient and robust MPT even in a sparse scattering environment. Third, the UA can support multiuser communications so long as they have sufficiently large separation distances as discussed in the sequel. In contract, the super-directive array is incapable of decoupling users with small separation in angle-of-arrival regardless of their separation distances. Last, the UA enables blind channel estimation while the super-directive array requires training based channel estimation as discussed later.

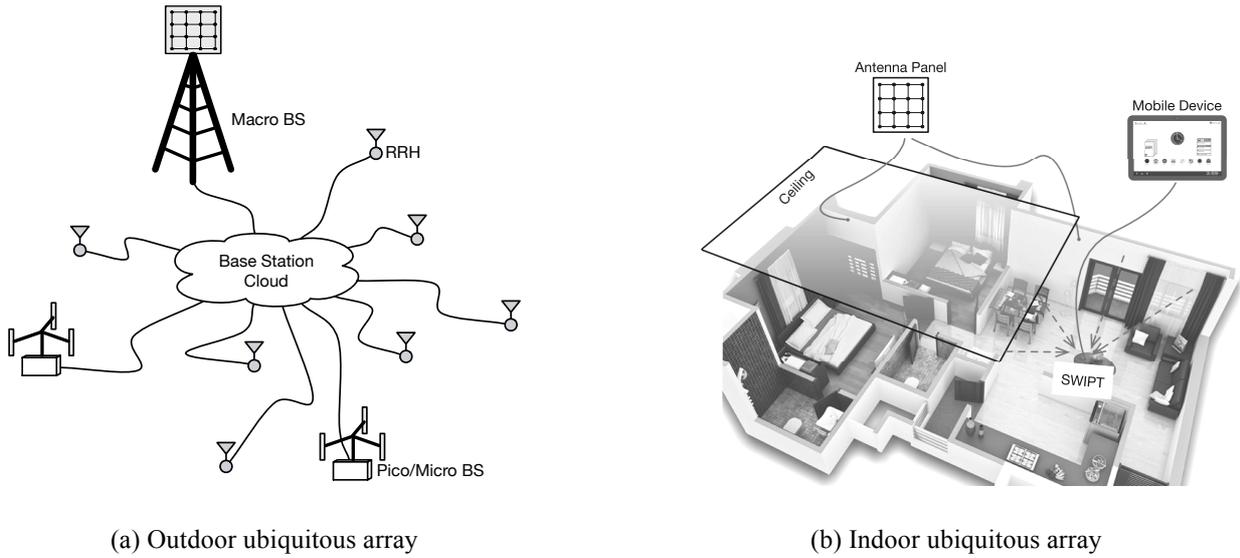

(a) Outdoor ubiquitous array                          (b) Indoor ubiquitous array

Fig. 4. The realization of a UA in (a) an outdoor environment via cloud based coordination of remote radio heads and heterogeneous BSs, and (b) an indoor environment by coordinating antenna panels embedded in walls or ceilings.

The performance of multiuser IT using the UA depends on the separation distances between users. If users are relatively far away from each other (e.g., tens of wavelengths), it is sufficient to form multiple modulated energy hotspots at corresponding users. The sharp decay of field power density with an increasing distance from the peak of a hotspot ensures small inter-user interference. It is shown in [15] that the decay rate for a dense spherical UA is the inverse of squared distance (measured in wavelength). If users are relatively close to each other, interference avoidance is necessary. The number of spatial DoF resulting from exciting phase modes of the UA is found to be approximately *proportional to the square of minimum user separation distance (in wavelength)*. An attempt in exciting more modes can lead to dramatic increase in transmission power to ensure required receive SNRs.

With ubiquitous and dense elements, a UA enables *blind* (without training) estimation/tracking of mobile locations, which is essential for forming the mentioned energy hotspots and mitigating interference. A relevant technique proposed in [15] exploits the *spatial signature* of a mobile resulting from the propagation phase shits for the large number of links between UA elements and mobile antennas. Given the information on UA antenna positions, the technique computes the modulus of the inner product between the sequence of observations by UA antennas and each of a set of spatial signatures generated for a set of random hypothesized locations. The result is a positive function of the hypothesized locations, called a *channel observation profile*, with peaks near/at the mobile locations. It follows that detecting the peaks of the profile yields the estimated mobile locations. Since no channel training is required, the BS can continuously update the profile to track the mobile movements for ensuring efficient SWIPT.

**b) Research Opportunities**

The deployment of UAs for SWIPT promises substantial gain for the PT efficiency and IT rate. However, due to the new propagation model for UA transmission, the design and analysis of SWIPT systems with UAs require the interplay between stochastic geometry, propagation theory and communication/information theory. This introduces a set of interesting research opportunities with several described as follows.

- **SWIPT using UAs over sparse scattering channels.** A UA channel can contain sparse scatterers that block the lines-of-sight between a subset of UA antennas and mobile devices. As a result, scattering reduces the PT efficiency, affects the spatial DoF for IT and compromises the accuracy of tracking device locations using the technique mentioned earlier. This makes it important to quantify the effects of sparse scattering on the SWIPT performance and the analysis that should account for the geometry, density and spatial distributions of scatterers.

- **SWIPT using randomly distributed UA antennas.** In practice, the UA antennas are randomly placed in the environment such as walls, roofs and lamp-poles instead of being configured to form an array with regular geometry e.g., circle or sphere as assumed in [15] for tractability. One promising approach is to model UA antennas as a spatial point processes in addition to other processes modelling scatterers and mobile devices, and then apply both the stochastic geometry and wave propagation theories to analyze the performance of the SWIPT network equipped with UAs.

- **Energy-rate tradeoff for SWIPT using UAs.** On one hand, from the perspective of improving the network-wise PT efficiency, it is desirable to reduce user separation distances and antenna spacing at mobiles. On the other hand, from the perspective of increasing the IT sum rate, the opposite is preferred, which suppresses inter-user interference and facilitates spatial multiplexing. This leads to an energy-rate tradeoff regulated by location based user scheduling and mobile antenna spacing. It is interesting to quantify the contribution of optimizing this tradeoff to the performance improvement for a SWIPT network.

## VI. Analog Spatial Decoupling for PB-Assisted WPC

a) **PT-IT Near-Far Problem in the PB-assisted WPC Systems**

To explain the PT-IT near-far problem, consider the PB-assisted WPC system in Fig. 5(a) that one PB (with a backhaul connection) and one BS perform SWIPT and IT to a single mobile, respectively. The SWIPT and IT signals carry independent data streams but intended for the same user. The user attempts to retrieve all the information streams in addition to harvesting energy from the SWIPT signal. The short-range SWIPT and long-range IT channels are suitably modelled correspondingly as free-space and Rayleigh fading (rich scattering) channels. Free-space beamforming for SWIPT, essential for high MPT efficiency, is possible due to sparse scattering in the short-range channel between the PB and mobile. The incident signal at the user is split such that almost all received energy is harvested and a negligible fraction is used for information decoding since its performance depends only on the SNR but not the power itself.

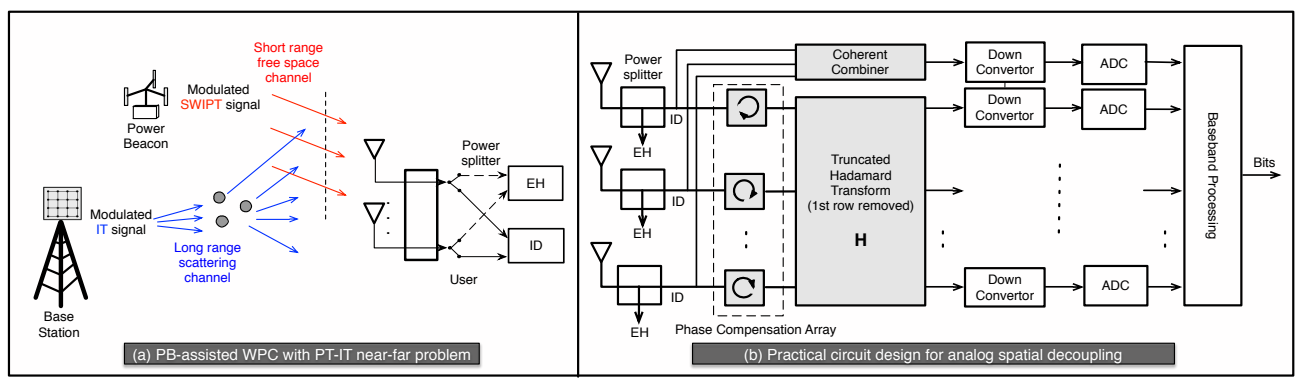

Fig. 5. (a) A model for PB-assisted WPC with PT-IT near-far problem, and (b) a practical receiver structure for analog spatial decoupling.

In this system, the **PT-IT near-far problem** arises from quantizing the mixture of strong SWIPT and weak IT signals. Their difference in power ranges from tens of dB for powering small devices such as sensors to more than 100 dB for powering large devices such as tablet computers [3]. In the quantization process illustrated in Fig. 6, the SWIPT signal is scaled to span the full dynamic range of the *analog-to-digital* *convertor* (ADC), which can reduce the peak magnitude of IT signal to be smaller than the quantization step size. This leads to extremely small *signal-to-quantization-noise ratio* (SQNR) for the IT signal after quantization. For example, given the power ratio of SWIPT and IT signals being 90 dB and a 10-bits ADC, the SQNR can be computed to be approximately -30 dB which is way too low for decoding the data streams in the IT signal. Therefore, for retrieving information from the IT signal in the presence of strong SWIPT signal, it is essential to decouple them in the analog domain and quantize them separately.

Time division and frequency division PT-IT are the conventional methods for overcoming the near-far problem but not without drawbacks. Time switching PT and IT reduce their efficiency/rate and further more requires synchronization between users. For frequency division PT-IT, the suppression of ultra-strong PT signal at an information decoder requires a sharp analog band-pass filter plus sufficiently large frequency separation between PT and IT signals. SWIPT using the same spectrum does not have the drawbacks mentioned above but requires *analog spatial decoupling* of IT and PT to solve their near-far problem, which is discussed in detail in the following sub-section.

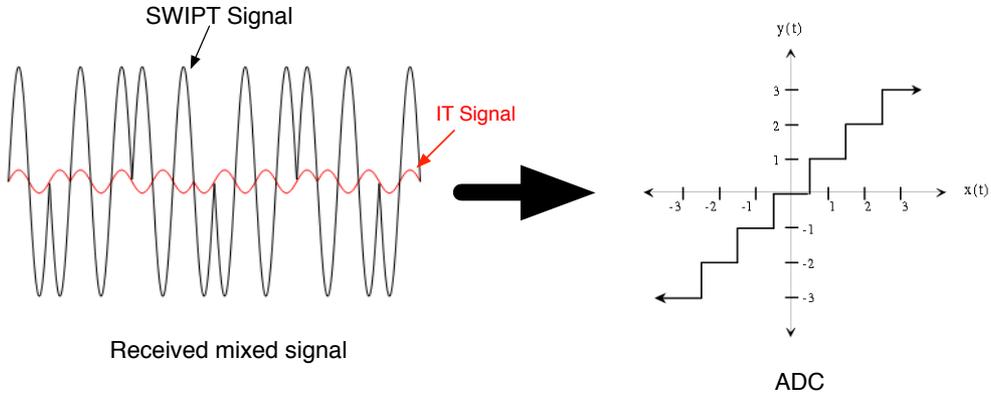

Fig. 6. Quantization of the received mixed strong-and-weak signals.

**b) Analog Spatial Decoupling**

Digital spatial decoupling of vector signals received from an antenna array is straightforward and involves implementation of the solution for a set of linear equations using a DSP processor. Analog spatial decoupling adds an additional analog circuit to the receiver and it is desirable to design it using simple components so as not to significantly increase the complexity and power consumption of the mobile device. The design constraints on analog decoupling make it difficult to extend the conventional digital approach and call for a novel and low complexity approach. Based on the system in Fig. 5 (a), a simple design using only phase shifters and adders as proposed in [12] is illustrated in Fig. 5 (b) and described as follows. This analog design not only decouples the SWIPT and IT signals but also achieves the maximum spatial multiplexing gain equal to the number of receive antennas, $N$. Specifically, The PB can transmit only a single data stream over the free-space channel that is denoted as $d_1(t)$. The data stream encounters different propagation phase rotations when it arrives at different receive antennas at the user. These phase rotations are compensated by the *phase compensation array* in Fig. 5 (b), giving the output $\rho d_1(t)[1, 1, \ldots, 1]^T$, where the scalar $\rho$ captures the path loss. The next component of the design, *truncated Hadamard transform*, multiplies this output with a truncated Hadamard matrix with the first row removed. For example, the truncated Hadamard matrix for the case of 4 receive antennas is shown as follows:

$$\begin{bmatrix} 1 & -1 & 1 & -1 \\ 1 & 1 & -1 & -1 \\ 1 & -1 & -1 & 1 \end{bmatrix}$$

Since the rows of the Hadamard matrix are orthogonal, the multiplication suppresses the SWIPT signal $d_1(t)$. The matrix comprises only -1 and 1 elements, which allows low-complexity implementation using $180^o$ phase shifters and adders. Next, consider the IT signal at the output of the truncated Hadamard transform. Note that the truncated Hadamard matrix has the full row rank and the Rayleigh MIMO channel matrix between the BS and user has an isotropic distribution invariant to the phase rotations imposed by the Phase Compensation Array. Therefore, assuming the BS has more than N antennas, the effective channel combining the said MIMO channel and the two analog operations has a rank equal to ($N$-1) and thus is capable of supporting ($N$-1) data streams transmitted by the BS. The quantization of these streams is free of the near-far problem since the SWIPT signal is suppressed by the analog operations. In addition, the data stream $d_1(t)$ transmitted by the PB is unaffected by the near-far problem and can be thus retrieved without any analog processing as shown in Fig. 5 (b). In summary, the low-complexity analog decoupling design presented in Fig. 5 (b) solves the IT-PT near far problem without sacrificing the spatial multiplexing gain.

c) **Research Opportunities**

Analog spatial decoupling is a new concept in the field of WPC and points to numerous new research opportunities with some described as follows.

- **Robust analog spatial decoupling**. Inaccurate channel estimation and non-ideal hardware can cause errors in the phase shifts of the decoupling circuit. This leads to a residual SWIPT signal mixed with the IT signal during its quantization, reducing the SQNR of the latter. Thus, it is important to characterize the effects of the practical factors on the performance of a PB assisted WPC system. Furthermore, new techniques can be designed to improve the robustness of analog spatial decoupling.

- **Tradeoff between training and transmission**. It follows from the above discussion that the performance of the PB-assisted WPC system would benefit from a longer channel estimation period yielding more accurate CSI estimate, but also suffer from a shorter energy/information transmission time. Therefore, how to achieve the optimal balance between channel estimation and the energy/information transmission is a very interesting topic deserving further investigation.

- **Analog spatial decoupling with multiple PBs**. The proposed receiver structure can only deal with the single PB case. When considering a more general multiple PBs case, it is challenging to design a feasible analog decoupling circuit. The major obstacle for solving the general case is to satisfy the unit-modulus constraints introduced by phase shifters. Mathematically, the design problem is equivalent to finding out a *SWIPT-signal-zero-forcing* matrix, whose rank is equal to the dimension of the null space of SWIPT signals and whose entries must have unit modulus. It is an interesting open problem whose solution may require some advanced mathematical tools from e.g., Fourier transform or manifold signal processing.

- **Analog decoupling with decision feedback**. Extending the classic approach of successive interference cancelation, an alternative design for analog decoupling is to first detect the strong SWIPT data stream in the digital domain and then feedback the decisions to cancel the SWIPT signal in the analog domain prior to detecting the IT signal. The main challenge for this approach is that the close-loop delay causes symbol-level misalignment between the received signal and feedback signal, rendering the cancelation ineffectiveness. Since the cancellation is in the analog domain, reversing the delay by DSP is infeasible and a novel design is required.

# VII. Concluding Remarks

Designing an efficient WPC system is much more complex than simply overlaying a communication and a MPT systems. The seamless PT-and-IT integration requires redesigning system and network architectures, techniques for transmissions and resource allocation, protocols for medium access control, and transceiver hardware. This has created a wide-range of research opportunities that drive the fast expansion of the field of WPC. In this article, we have discussed several interesting new research trends that can play a key role in transforming WPC from theory to practice. They include the use of backscatter antennas to realize WPC for low-complexity devices, the waveform design for improving the MPT efficiency and analog decoupling of the PT-and-IT to solve their near-far problem. The limit of SWIPT performance achievable by sharp beamforming is a fundamental issue. In this article, we have also discussed how to quantify this limit by leveraging recent theoretic results on super-directivity and how to improve the limit by deploying ubiquitous arrays. This article together with others in the current special issue will provide an updated roadmap of the WPC area and suggest diversified and promising directions for advancing its frontier.

# Acknowledgment

The work of Caijun Zhong was supported by the Zhejiang Provincial Natural Science Foundation of China (LR15F010001).

# Biographies

Kaibin Huang [S'05, M'08, SM'13] (huangkb@eee.hku.hk) received his Ph.D. degree from the University of Texas at Austin in electrical engineering. Since January 2014, he has been an assistant professor in the Department of EEE at the University of Hong Kong. He is an editor for IEEE JSAC – Series on Green Communications and Networking, IEEE Transactions on Wireless Communications and also IEEE Wireless Communications Letters. He received a Best Paper Award from IEEE GLOBECOM 2006 and an IEEE Communications Society Asia Pacific Outstanding Paper Award in 2015. His research interests focus on the analysis and design of wireless networks using stochastic geometry and multi-antenna techniques.

Caijun Zhong [S'07, M'10, SM'14] (caijunzhong@zju.edu.cn) received his Ph.D. degree in electrical and electronic engineering from University College London in 2010. He is currently an associate professor at Zhejiang University. His research interests include Massive MIMO, full-duplex relaying, and wireless powered communications. He serves as an editor for several journals including IEEE Transactions on Wireless Communications, IEEE Communications Letters, and EURASIP Journal on Wireless Communications and Networking.

Guangxu Zhu [S'14] (gxzhu@eee.hku.hk) received his B.S. and M.S. degree in Information and Communication Engineering from the Zhejiang University in 2012 and 2015, respectively. He is currently working towards his Ph.D. degree in the the department of EEE at the University of Hong Kong. His research interests include MIMO communications systems, cooperative communications and wirelessly powered communications. He is the recipient of a Best Paper Award from WCSP 2013.